\documentclass[aps,prl,twocolumn,superscriptaddress,groupedaddress,preprintnumbers]{revtex4} 
\usepackage{graphicx}  
\usepackage{dcolumn}   
\usepackage{bm}        
\usepackage{amssymb}   
\usepackage{amsmath}
\usepackage{mathtools}
\usepackage{cases}
\usepackage{mathrsfs}
\usepackage{color}
\usepackage{CJK}
\usepackage{ulem}

\newcommand{\be}{\begin{equation}}
\newcommand{\ee}{\end{equation}}
\newcommand{\bea}{\begin{eqnarray}}
\newcommand{\eea}{\end{eqnarray}}
\newcommand{\nn}{\nonumber}
\newcommand{\bx}{\mathbf{x}}
\newcommand{\bk}{\mathbf{k}}

\newcommand{\bp}{\mathbf{p}}

\newcommand{\fnl}{F_\text{NL}}
\begin{document}
\begin{CJK*}{GB}{}
	\preprint{IPMU18-0172}
	\preprint{YITP-18-114}
	\title{Gravitational Waves Induced by Non-Gaussian Scalar Perturbations}
	\author{Rong-Gen Cai${}^{a,b}$,~Shi Pi${}^c$ and Misao Sasaki${}^{c,a,d,e}$\\
		\it
		$^{a}$CAS Key Laboratory of Theoretical Physics, \\
		Institute of Theoretical Physics,
		Chinese Academy of Sciences, Beijing 100190, China\\
		$^{b}$School of Physical Sciences,
		University of Chinese Academy of Sciences, Beijing 100049, China\\
		$^{c}$Kavli Institute for the Physics and Mathematics of the Universe (WPI), Chiba 277-8583, Japan\\
		$^{d}$Yukawa Institute for Theoretical Physics, Kyoto University, Kyoto 606-8502, Japan\\
		$^{e}$Leung Center for Cosmology and Particle Astrophysics,\\National Taiwan University, Taipei 10617}
	    
	\date{\today}
	\begin{abstract}
We study gravitational waves (GWs) induced by non-Gaussian curvature perturbations. 
We calculate the density parameter per logarithmic frequency interval,  $\Omega_\text{GW}(k)$, 
given that the power spectrum of the curvature perturbation
 $\mathcal{P}_\mathcal{R}(k)$ has a narrow peak at some small scale $k_*$, with a local-type non-Gaussianity, 
 and constrain the nonlinear parameter $f_\text{NL}$ with the future LISA sensitivity curve as well as 
 with constraints from the abundance of the primordial black holes (PBHs). 
 We find that the non-Gaussian contribution to $\Omega_\text{GW}$ increases as $k^3$, peaks at $k/k_*=4/\sqrt{3}$,
 and has a sharp cutoff at $k=4k_*$. The non-Gaussian part can exceed the Gaussian part if $\mathcal{P}_\mathcal{R}(k)f_\text{NL}^2\gtrsim1$. If both a slope $\Omega_\text{GW}(k)\propto k^\beta$ with $\beta\sim3$ and the multiple-peak structure around a cutoff are observed,
 it can be recognized as a smoking gun of the primordial non-Gaussianity. 
  We also find that if PBHs with masses of $10^{20}$ to $10^{22}\text{g}$ are identified as cold dark 
  matter of the Universe, the corresponding GWs must be detectable by LISA-like detectors, irrespective of
  the value of $\mathcal{P}_\mathcal{R}$ or $f_\text{NL}$. 
\end{abstract}
	\maketitle
	\end{CJK*}

\textit{Introduction.}~
The detection of gravitational waves (GWs) from mergers of  black holes (BHs) or neutron stars (NSs) by LIGO/VIRGO~\cite{Abbott:2016blz,Abbott:2016nmj,Abbott:2017vtc,Abbott:2017gyy,Abbott:2017oio,TheLIGOScientific:2017qsa} has marked the beginning of the era of gravitational wave astronomy. Besides these GWs from mergers, there are other sources of GWs, like BH/NS binaries~\cite{Regimbau:2011rp,Zhu:2011bd,Rosado:2011kv,Marassi:2011si,Wu:2011ac,Zhu:2012xw,Wu:2013xfa,Mandic:2016lcn,Clesse:2016ajp,Wang:2016ana,Raidal:2017mfl}, phase transitions during the evolution of the universe~\cite{Kosowsky:1992vn,Kamionkowski:1993fg,Apreda:2001us,Grojean:2006bp,Caprini:2007xq,Caprini:2009yp,Hindmarsh:2013xza,Huang:2016odd,Cai:2017tmh,Chao:2017vrq,Caprini:2009fx,Jinno:2016vai,Cutting:2018tjt}, reheating or preheating after inflation~\cite{Tashiro:2003qp,Easther:2006vd,GarciaBellido:2007dg,Dufaux:2007pt,Kuroyanagi:2015esa,Liu:2017hua,Kuroyanagi:2017kfx,Cai:2018tuh}, and primordial scalar and tensor perturbations from inflation. For reviews of GW physics, see Refs.~\cite{Sathyaprakash:2009xs,Guzzetti:2016mkm,Cai:2017cbj}.


The amplitude of the primordial tensor perturbation is much smaller than that of the scalar curvature perturbation
 on CMB scales which is $\mathcal{R}\sim\mathcal{O}(10^{-5})$. The current constraint on the tensor-to-scalar ratio $r$ is $r<0.064$ at the $95\%$ level~\cite{Akrami:2018odb}. 
However, as the scalar and tensor perturbations are coupled at the nonlinear level, 
we do have an induced tensor perturbation of order $\mathcal R^2$.
In most of the inflation models the induced tensor perturbation is much smaller than 
the primordial one from the vacuum fluctuations. 
Nevertheless, there are models of inflation that predict large curvature perturbations on small scales~\cite{GarciaBellido:1996qt,Kawasaki:1997ju,Yokoyama:1998pt,Frampton:2010sw,Kawasaki:2012wr,Inomata:2017okj,Garcia-Bellido:2017mdw,Kannike:2017bxn,Inomata:2017vxo,Ando:2017veq,Ando:2018nge,Espinosa:2017sgp,Pi:2017gih,Espinosa:2018eve,Kohri:2012yw,Clesse:2015wea,Cheng:2016qzb,Cheng:2018yyr,Garcia-Bellido:2016dkw}, for which the induced tensor perturbation may dominate over the primordial one.

Early works on GWs induced by the scalar perturbation at second order can be found in Refs.~\cite{Matarrese:1992rp,Matarrese:1993zf,Matarrese:1997ay,Noh:2004bc,Carbone:2004iv,Nakamura:2004rm}. In Refs.~\cite{Ananda:2006af,Baumann:2007zm}, the evolution of the induced GWs in the radiation-dominated era was studied. It was found that a $\delta$-function-like peak in the power spectrum of the curvature perturbation $\mathcal{P}_\mathcal{R}\sim\delta(k-k_*)$ may induce a characteristic GW power spectrum, which has a zero point at $k/k_*=\sqrt{2/3}$, and a peak at $k/k_*=2/\sqrt{3}$. This behavior was then confirmed numerically and analytically in Refs.~\cite{Osano:2006ew,Alabidi:2012ex,Alabidi:2013wtp,Inomata:2016rbd,Orlofsky:2016vbd,Kohri:2018awv,Assadullahi:2009jc,Biagetti:2014asa,Gong:2017qlj,Giovannini:2010tk}.

Current CMB data do not exclude the possibility that the scalar perturbation is large on small scales~\cite{Bringmann:2011ut,Green:2018akb}. Typically, if the power spectrum for the primordial curvature
 perturbation has a peak on some small scale, there may be some regions where the density perturbation 
 exceeds a threshold value $\delta_\text{th}\sim0.3$ at horizon reentry, and the matter inside the Hubble horizon collapses to 
 form a primordial black hole (PBH) ~\cite{Zeldovich:1963,Hawking:1971ei,Carr:1974nx,Meszaros:1974tb,Carr:1975qj}. 
The mass of a PBH is of the same order of the total energy inside the Hubble radius at horizon reentry,
which is hence determined by the wavenumber of the peak. Various constraints on the abundance of PBHs have been
discussed~\cite{Frampton:2009nx,Frampton:2010sw,Carr:2009jm,Carr:2016drx,Green:2004wb,Bird:2016dcv,Pi:2017gih,
	Regimbau:2011rp,Zhu:2011bd,Rosado:2011kv,Marassi:2011si,Wu:2011ac,Zhu:2012xw,Wu:2013xfa,Mandic:2016lcn,Clesse:2016ajp,Clesse:2016vqa,Sasaki:2016jop,Chen:2016pud,Blinnikov:2016bxu,Ali-Haimoud:2016mbv,Garcia-Bellido:2017aan,Zumalacarregui:2017qqd,Garcia-Bellido:2017imq,Kawasaki:2012wr,Guo:2017njn}. 

The relation between the induced GWs and PBH formation was first studied in Refs.~\cite{Saito:2008jc}
 for a $\delta$-function peak of power spectrum, and then for broad plateaus  by Refs.~\cite{Saito:2009jt,Bugaev:2009zh,Bugaev:2010bb}. 
However, in those previous studies, the scalar perturbation was assumed to be Gaussian,
which seems to be a rather naive assumption. When there appears a sharp peak in the curvature
perturbation spectrum, it is natural to expect that there also appears a non-negligible non-Gaussianity.
As PBHs are produced at the large amplitude tail of the probability distribution of the curvature perturbation,
any non-negligible non-Guassianity would completely alter the PBH formation rate. This also suggests
that we may have very different predictions on the amplitude and shape of the induced GW spectrum.
GWs induced by the non-Gaussian scalar perturbation were estimated by Refs.~\cite{Nakama:2016gzw}, while the large non-Gaussianity limit in a concrete model was studied in \cite{Garcia-Bellido:2016dkw}. In this Letter we study the GWs induced by non-Gaussian scalar perturbation in general.

\textit{Induced gravitational waves.}~
The perturbed metric in the Newton gauge is
\be\nn
ds^2=a^2\!\!\left[-\left(1-2\Phi\right)d\eta^2\!+\!\left((1+2\Phi)\delta_{ij}+h_{ij}\!\right)\!dx^idx^j\right].
\ee
where $\eta$ is the conformal time, $\Phi$ is the curvature perturbation in the Newton gauge, and $h_{ij}$ is the tensor perturbation,
and we have neglected the anisotropic stress perturbation~\cite{Weinberg:2003ur,Watanabe:2006qe,Baumann:2007zm}.
The equation for the Fourier component of the tensor perturbation at second order 
in the radiation dominated universe reads~\cite{Ananda:2006af}
\bea\nn
&&h_\bk''+2\mathcal{H}h_\bk'+k^2h_\bk\\\nn
&=&18\int\frac{d^3l}{(2\pi)^{3/2}}\frac{l^2}{\sqrt2}\sin^2\theta
\left(
\begin{matrix}
\cos2\varphi\\
\sin2\varphi
\end{matrix}
\right)
\Phi_\mathbf{l}\Phi_{\bk-\mathbf{l}}\\\nn
&&\;\;\;\;\times\left[j_0(ux)j_0(vx)-2\frac{j_1(ux)j_0(vx)}{ux}\right.\\\label{S2}
&&\;\;\;\;\;\;\;\;
\left.-2\frac{j_0(ux)j_1(vx)}{vx}+6\frac{j_1(ux)j_1(vx)}{uvx^2}\right].\label{eom:h}
\eea
Here, $\cos2\varphi$ or $\sin2\varphi$ is for $+$ or $\times$ polarization. 
We also define new variables  $u=|\bk-\mathbf{l}|/k$,  $v=l/k$ and $x=k\eta/\sqrt3$. 
Equation \eqref{eom:h} can be solved by the Green function method. 
After solving $h_\bk$, we can use its two-point correlation function to calculate the density parameter
 $\Omega_\text{GW}(k)$ defined as the energy density of the GW per unit logarithmic frequency normalized 
by the critical density, 
\be\label{def:OmegaGW}
\Omega_\text{GW}(k)\equiv\frac1{12}\left(\frac{k}{Ha}\right)^2\frac{k^3}{\pi^2}\overline{\langle h_\bk(\eta)h_\bk(\eta)\rangle}, 
\ee
where the overline means the time average. 
It then follows that because the contribution from the connected four-point function vanishes by symmetry, the two-point function of $h_\bk$ can be deduced to a product of the two-point functions of $\Phi_\bk$'s.
For convenience, we change the variable to the curvature perturbation in comoving slices $\mathcal{R}$, 
which is related to $\Phi$ by $\Phi=(2/3)\mathcal{R}$ on superhorizon scales
 in the radiation-dominated universe.  Up to the second order, it is expressed in terms of the Gaussian part
 as~\cite{Luo:1992er,Verde:1999ij,Verde:2000vr,Komatsu:2001rj,Bartolo:2004if,Boubekeur:2005fj,Byrnes:2007tm}: 
\be\label{def:NGR}
\mathcal{R}(\bx)=\mathcal{R}_g(\bx)+F_\mathrm{NL}\left[\mathcal{R}_g^2(\bx)-\langle\mathcal{R}_g^2(\bx)\rangle\right],
\ee
where we have introduced the nonlinear parameter for $\mathcal{R}$, $\fnl$, which is related to the nonlinear
 parameter for $\Phi$, $f_\text{NL}$, by $F_\text{NL}=(3/5)f_\text{NL}$. 
Then for the two-point correlation function of $\Phi_\bk$, we have
\be\label{psipsi}
\langle\Phi_{\bk}\Phi_{\mathbf{p}}\rangle\sim\frac49\left(P_\mathcal{R}(k)+2\fnl^2\int d^3l~P_\mathcal{R}(|\bk-\mathbf{l}|)P_\mathcal{R}(l)\right),
\ee
where we omitted an overall factor $(2\pi)^3\delta^{(3)}(\bk+\mathbf{p})$, and the Gaussian power spectrum is defined as $\langle\mathcal{R}_{g,\bk}\mathcal{R}_{g,\mathbf{p}}\rangle=(2\pi)^3P_\mathcal{R}(k)\delta^{(3)}(\bk+\bp)$. 

To step forward, we should specify the $k$ dependence of $P_\mathcal{R}(k)$, which in general can be 
different from the nearly scale-invariant spectrum we observe on the CMB scales. 
Here we study the case of a primordial curvature perturbation with a narrow peak at some specific scale $k_*$ 
with a width $\sigma\ll k_*$,
\be\label{def:PR}
P_\mathcal{R}(k)=\frac{\mathcal{A}_\mathcal{R}}{(2\pi)^{3/2}2\sigma k_*^2}\exp\left(-\frac{(k-k_*)^2}{2\sigma^2}\right).
\ee
The coefficient is to normalize $\int d^3kP_\mathcal{R}(k)=\mathcal A_\mathcal{R}$. 
This power spectrum with a narrow peak can be produced in various models of 
inflation~\cite{Frampton:2010sw,Kawasaki:2012wr,Pi:2017gih}, and easy to be extended to more general cases. 
We neglect the scale invariant contribution extrapolated from the CMB scales, since we assume $\mathcal{A}_\mathcal{R}$ 
is much larger than $10^{-9}$. Keeping in mind that $\sigma\ll k_*$, we can
 calculate the convolution of the power spectra in Eq.~\eqref{psipsi},
\begin{align}\label{result:intPP}
\int d^3l~P_\mathcal{R}(|\bk-\mathbf{l}|)P_\mathcal{R}(l)
\approx\frac{\mathcal{A}_\mathcal{R}^2}{(2\pi)^2}\frac{\pi}{2kk_*^2}\text{erf}\left(\frac{k}{2\sigma}\right),
\end{align}
where 
terms suppressed by higher orders of $\sigma/k_*$ are neglected. 
When $k>2k_*$, there is an exponentially suppressed tail which we can safely neglect. 
Then we can calculate the power spectrum of the tensor perturbation, up to the epoch of radiation-matter equality. 
Using Eq.~\eqref{def:OmegaGW}, we obtain
\begin{align}\nn
\Omega_\text{GW}
&=6\mathcal A_\mathcal{R}^2\frac{k^2}{2\pi\sigma^2}\left(\frac{k}{k_*}\right)^4
\int^\infty_0dv\int^{1+v}_{|1-v|}du\,uv\,\mathcal{T}(u,v)\\\nn
&\times\left[e^{-\frac{(vk-k_*)^2}{2\sigma^2}}
+2\mathcal A_\mathcal{R}\fnl^2\frac{\sigma}{vk}\sqrt{\frac\pi2}
\text{erf}\left(\frac{vk}{2\sigma}\right)\right]\\\label{OmegaFull}
&\times\left[e^{-\frac{(uk-k_*)^2}{2\sigma^2}}
+2\mathcal A_\mathcal{R}\fnl^2\frac{\sigma}{uk}\sqrt{\frac\pi2}
\text{erf}\left(\frac{uk}{2\sigma}\right)\right].
\end{align}
where the integral kernel, $\mathcal T(u,v)$, was derived by \cite{Kohri:2018awv}
\begin{align}
\mathcal{T}(u,v)
&=\frac14\left(\frac{4v^2-(1+v^2-u^2)^2}{4uv}\right)^2\left(\frac{u^2+v^2-3}{2uv}\right)^2\nonumber\\
&\times\left\{\left(-2+\frac{u^2+v^2-3}{2uv}\ln\left|\frac{3-(u+v)^2}{3-(u-v)^2}\right|\right)^2\right.\nonumber\\\label{def:T}
&+\left.\pi^2\left(\frac{u^2+v^2-3}{2uv}\right)^2\Theta\left(u+v-\sqrt3\right) \right\}.
\end{align}
If the non-Gaussian contribution is small, the leading order is given by the Gaussian integral. 
For a $\delta$-function-like peak of the curvature perturbation, the main contribution comes from the neighborhood of $u\sim v\sim k_*/k$,
which gives
\be\label{Omega(0)}
\Omega_\text{GW}^{(0)}\simeq6\mathcal{A}_\mathcal{R}^2\left(\frac{k}{k_*}\right)^2\mathcal{T}\left(\frac{k_*}{k},\frac{k_*}k\right)\Theta(2k_*-k).
\ee
When $k\ll k_*$, the leading term of $\mathcal{T}(k_*/k,k_*/k)$ is approximately a constant, so $\Omega_\text{GW}^{(0)}\propto k^2$, with a peak about $\Omega_\text{GW,peak}^\text{(0)}\simeq21.0\mathcal{A}_\mathcal{R}^2$ at $k_p^{(0)}\sim(2/\sqrt3)k_*$. When $k\ll\sigma\ll k_*$, $\Omega_\text{GW}^{(0)}\propto k^3$.
Detailed studies of this Gaussian case can be found in Refs.~\cite{Ananda:2006af,Saito:2008jc,Saito:2009jt,Alabidi:2012ex,Alabidi:2013wtp,Orlofsky:2016vbd,Kohri:2018awv}.

If $\mathcal{A}_\mathcal{R}\fnl^2\gtrsim\mathcal{O}(1)$, the contribution from the non-Gaussianity dominates 
the tensor power spectrum. The contributions from the terms proportional to $\fnl^2$ and
$\fnl^4$, respectively, have the form
\begin{align}\nn
\Omega_\text{GW}^{(2)}=&6\mathcal{A_R}^3\fnl^2\left(\frac{k}{k_*}\right)^3\Theta(3k_*-k)\\\nn
\times&\left[\int^{\text{min}(1+k_*/k,2k_*/k)}_{|1-k_*/k|}du~\mathcal{T}\left(u,\frac{k_*}{k}\right)\right.\\
&\left.+\int^{\text{min}(2k_*/k,1+k_*/k)}_{\text{max}(0,|k_*/k-1|)}
dv~\mathcal{T}\left(\frac{k_*}{k},v\right)\right].
\label{Omega2}\\\nn
\Omega_\text{GW}^{(4)}=&6\mathcal{A_R}^4\fnl^4\left(\frac{k}{k_*}\right)^4\int^\infty_0dv\int^{1+v}_{|1-v|}du\\
\times&\mathcal{T}(u,v)\Theta\left(2k_*-vk\right)\Theta\left(2k_*-uk\right).\label{Omega4}
\end{align}
If $\mathcal{A}_\mathcal{R}\fnl^2\gg1$, the term proportional to $\fnl^4$, $\Omega_\text{GW}^{(4)}$, 
overwhelms the terms proportional to $\fnl^2$, $\Omega_\text{GW}^{(2)}$.
It is nonzero only for $k<4k_*$, which is twice of the Gaussian cutoff at $2k_*$. 
It also has a peak at twice the frequency of the Gaussian peak, i.e. $k_p^{(4)}\sim(4/\sqrt3)k_*$.
 The scaling law when $k\ll k_*$ can be estimated by requiring $u\sim v\gg1$ in Eq.~\eqref{Omega4},
\begin{align}\label{Omega(4)}
\Omega_\text{GW}^\text{(4)}&\simeq89.6\left(\mathcal A_\mathcal{R}\fnl\right)^4\left(\frac{k}{k_*}\right)^3
\biggl[1+\cdots\biggr]\Theta(4k_*-k).
\end{align}
The dots represent terms proportional to $\ln(k/k_*)$ and $[\ln(k/k_*)]^2$ 
which can be neglected in the LISA sensitivity band. We can see it increases as $k^3$ when $k$ is small, 
which is faster than $k^2$. The GWs induced by non-Gaussian scalar purturbations are easily
distinguishable if they dominate, which depends on the ratio of the peak amplitudes, 
\be
\frac{\Omega_\text{GW}^\text{(4)}}{\Omega_\text{GW}^{(0)}}\sim4.3\mathcal{A}_\mathcal{R}^2\fnl^4.
\ee
If this is larger than unity, we will clearly see the effect of the non-Gaussianity.
For the power spectrum of $\mathcal{R}$,  $\fnl\gtrsim10$ will be enough for a peak amplitude of 
$\mathcal{A}_\mathcal{R}\sim10^{-2}$.
  We emphasize that there is no observational constraint on $\fnl$ on small scales. If $\mathcal{A}_\mathcal{R}^2\fnl^4\gtrsim1$, which implies 
that $\Omega_\text{GW}^{(4)}$ and $\Omega_\text{GW}^{(2)}$ are larger than $\Omega_\text{GW}^{(0)}$,
we find a series of peaks from the resonances around $k_p^{(0)}\sim(2/\sqrt3)k_*$, $k_p^{(2)}\sim\sqrt3k_*$, 
and $k_{p}^{(4)}\sim(4/\sqrt3)k_*$, which can be recognized as a smoking gun of the primordial non-Gaussianity 
at scale $k_*$. 

\begin{figure}
	\includegraphics[width=0.5\textwidth]{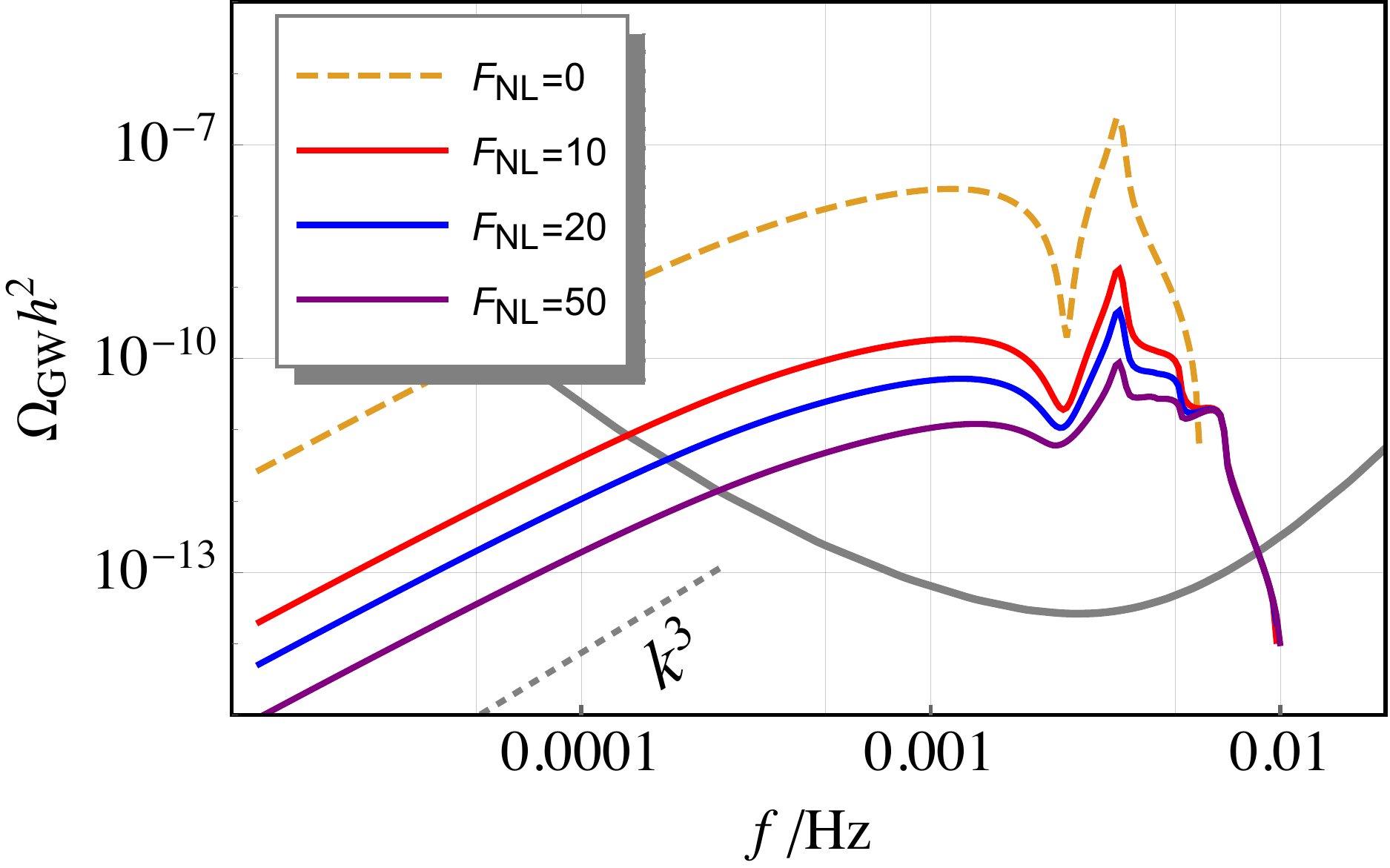}
	\\
	\vspace{7pt}
	\includegraphics[width=0.5\textwidth]{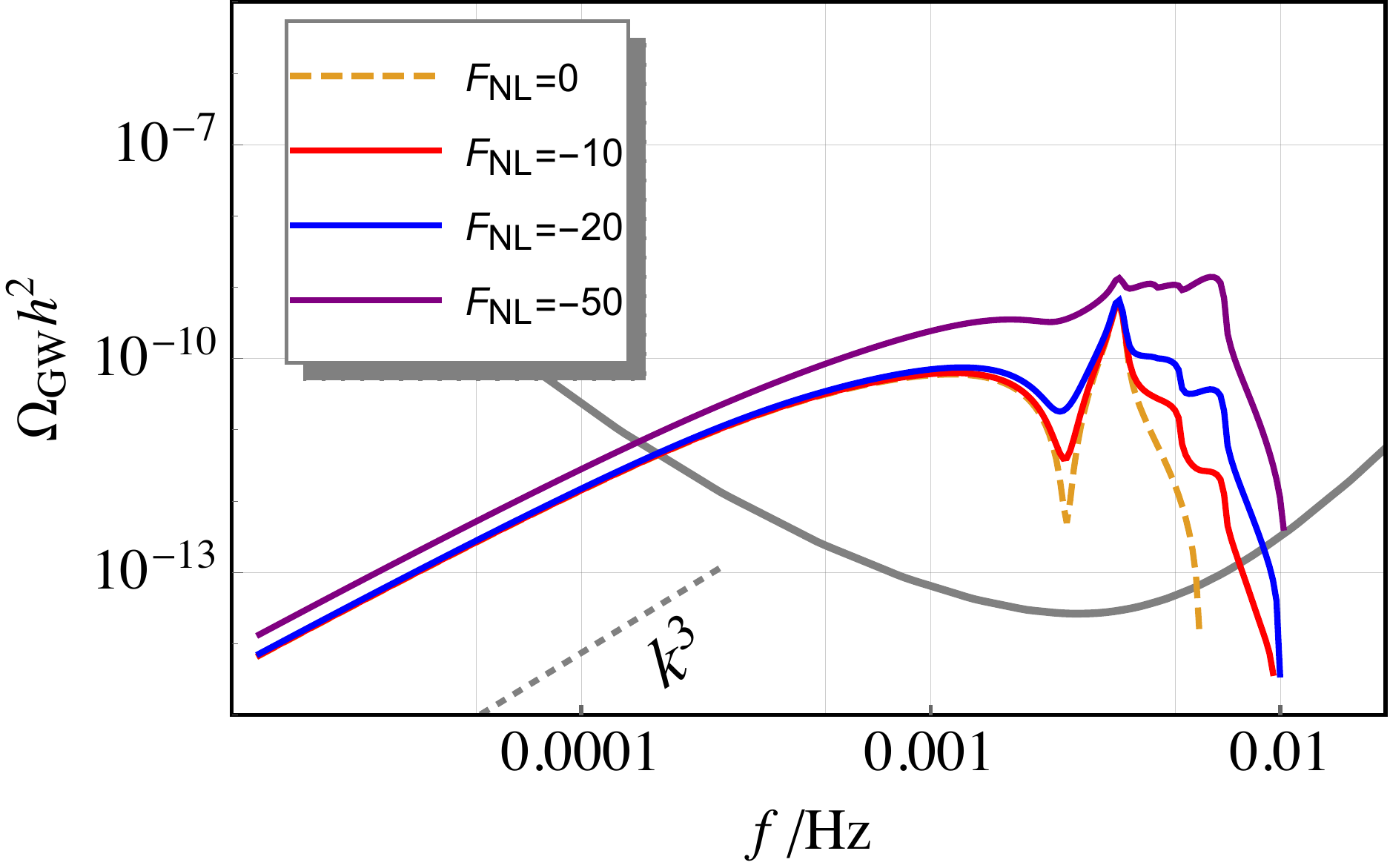}
	\caption{
		Typical gravitational wave density parameter induced by a non-Gaussian curvature perturbation at second order. 
			The width of peak is fixed at $\sigma=10^{-4}~\text{Hz}$. In the upper panel, $\fnl$ is positive, where the abundance of the PBHs is fixed to be $f_\text{PBH}=1$ for $M_\text{PBH}=10^{22}~\text g$. In the lower panel, $\fnl$ is negative, where the amplitude of the peak is fixed to be $\mathcal{A}_\mathcal{R}=10^{-3}$. In both cases, we have drawn $|\fnl|=0$ (orange dashed), $10$ (red), $20$ (blue), and $50$ (purple). The gray curve is the sensitivity bound of LISA from Ref.~\cite{Thrane:2013oya}. A reference line of the $k^3$ slope is also drawn for comparison.
		}\label{fig:detect}
\end{figure}

\textit{Observational implications.}~
The GW density parameter calculated in the previous section is valid from the horizon reentry to matter-radiation equality. The GW density parameter today is given by~\cite{Saito:2008jc,Saito:2009jt}
\be\nn
\Omega_\text{GW,0}h^2=4\times10^{-9}
\frac{\Omega_rh^2}{4\times10^{-5}}
\left(\frac{\mathcal{A}_\mathcal{R}}{10^{-2}}\right)^2\times\frac{\Omega_\text{GW,eq}}{\mathcal{A}_\mathcal{R}^2},
\ee
where $\Omega_\text{GW,eq}$ is the result obtained from Eq. \eqref{OmegaFull}, 
and we have neglected detailed dependence on the thermal history of the Universe studied in Refs.~\cite{Schwarz:1997gv,Weinberg:2003ur,Seto:2003kc,Boyle:2005se,Watanabe:2006qe,Kuroyanagi:2008ye,Saikawa:2018rcs} which may be easily incorporated if necessary. We see that the amplitude of the GW density parameter is determined by the peak value of the primordial scalar perturbation, which may generate PBHs whose masses are also determined by the frequency of the peak.
This was first studied by Saito \textit{et al}. in~\cite{Saito:2008jc}, 
\be\label{frequency-mass}
f_\text{GW}\sim3~\text{Hz}\left(\frac{M_\text{PBH}}{10^{16}\text{g}}\right)^{-1/2},
\ee
where $f_\text{GW}$ is related to $k$ by $f_\text{GW}=k/(2\pi a)$ where $a$ is the scale factor. We know that PBHs lighter than $5\times10^{14}~\text{g}$ have already evaporated by today through Hawking radiation, while PBHs lighter than $10^{16}~\text{g}$ are approaching their doomsday by radiating high energy particles which are strongly constrained by the observation of $\gamma$-ray background~\cite{Carr:2009jm}. 
This implies there is an upper bound for the frequency of GWs induced by scalar perturbations, $f_\text{GW}\lesssim3~\text{Hz}$. Therefore, unfortunately, we cannot expect any induced GWs 
to be detected by LIGO/VIRGO/KAGRA/ET ($10$ to $10^3~\text{Hz}$)~\cite{Aasi:2013wya,Punturo:2010zz,Sathyaprakash:2012jk}. 
However, we may see them by the next-generation GW observatories like LISA ($10^{-4}$ to $0.1$~Hz) \cite{AmaroSeoane:2012km,AmaroSeoane:2012je,Audley:2017drz}, Taiji~\cite{Guo:2018npi}, Tianqin~\cite{Luo:2015ght}, BBO ($0.1$ to 1~Hz)~\cite{Crowder:2005nr,Corbin:2005ny} 
or DECIGO ($10^{-2}$ to 1~Hz)~\cite{Kawamura:2006up,Kawamura:2011zz}. 
In Fig.~\ref{fig:detect}, the results of numerical integration of Eq. \eqref{OmegaFull} for different $\fnl$ and $\mathcal{A}_\mathcal{R}$ and the corresponding current density parameter $\Omega_\text{GW,0}h^2$
are shown, together with the LISA sensitivity curve.
As we can see, for a fixed $\mathcal{A}_\mathcal{R}$, smaller $\fnl$ will leave some resonance peaks 
as tails beyond the $2f_*$ peak, which may be difficult to detect. 
On the contrary, large $\fnl$ can make the resonance peaks prominent, while the peak around $(2/\sqrt3)f_*$ 
becomes barely visible. 

\begin{figure}
	\includegraphics[width=0.4\textwidth]{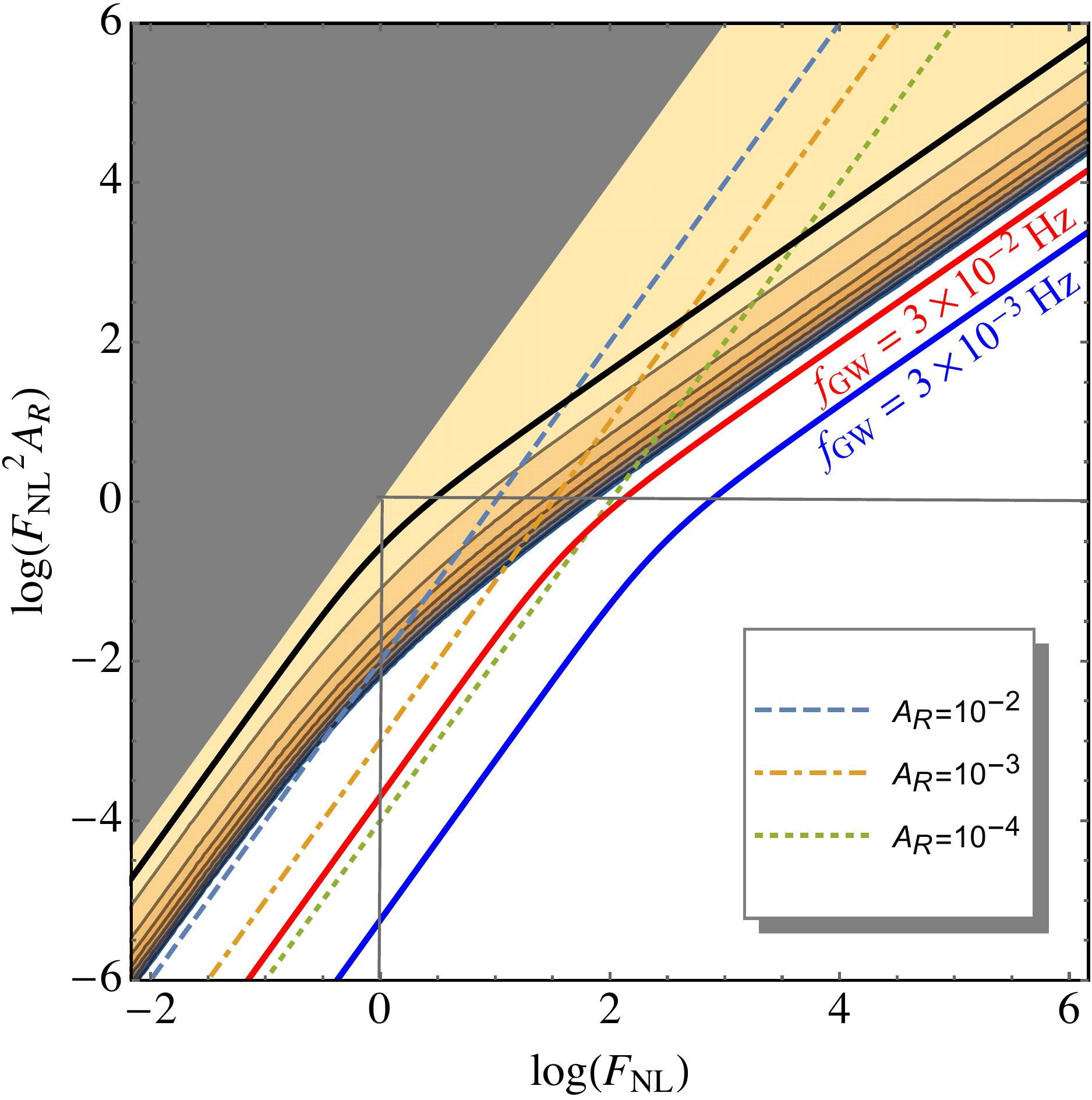}\includegraphics[width=0.05\textwidth]{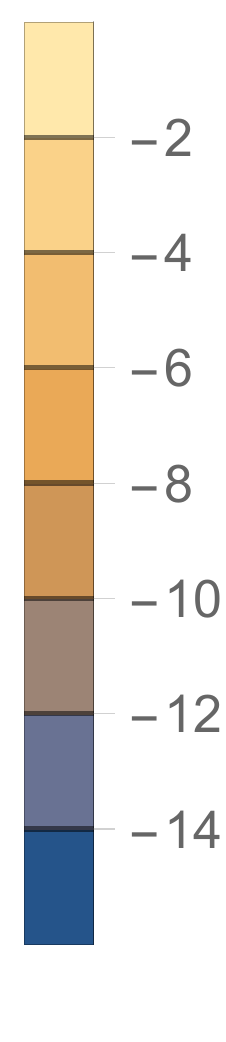}
	\includegraphics[width=0.4\textwidth]{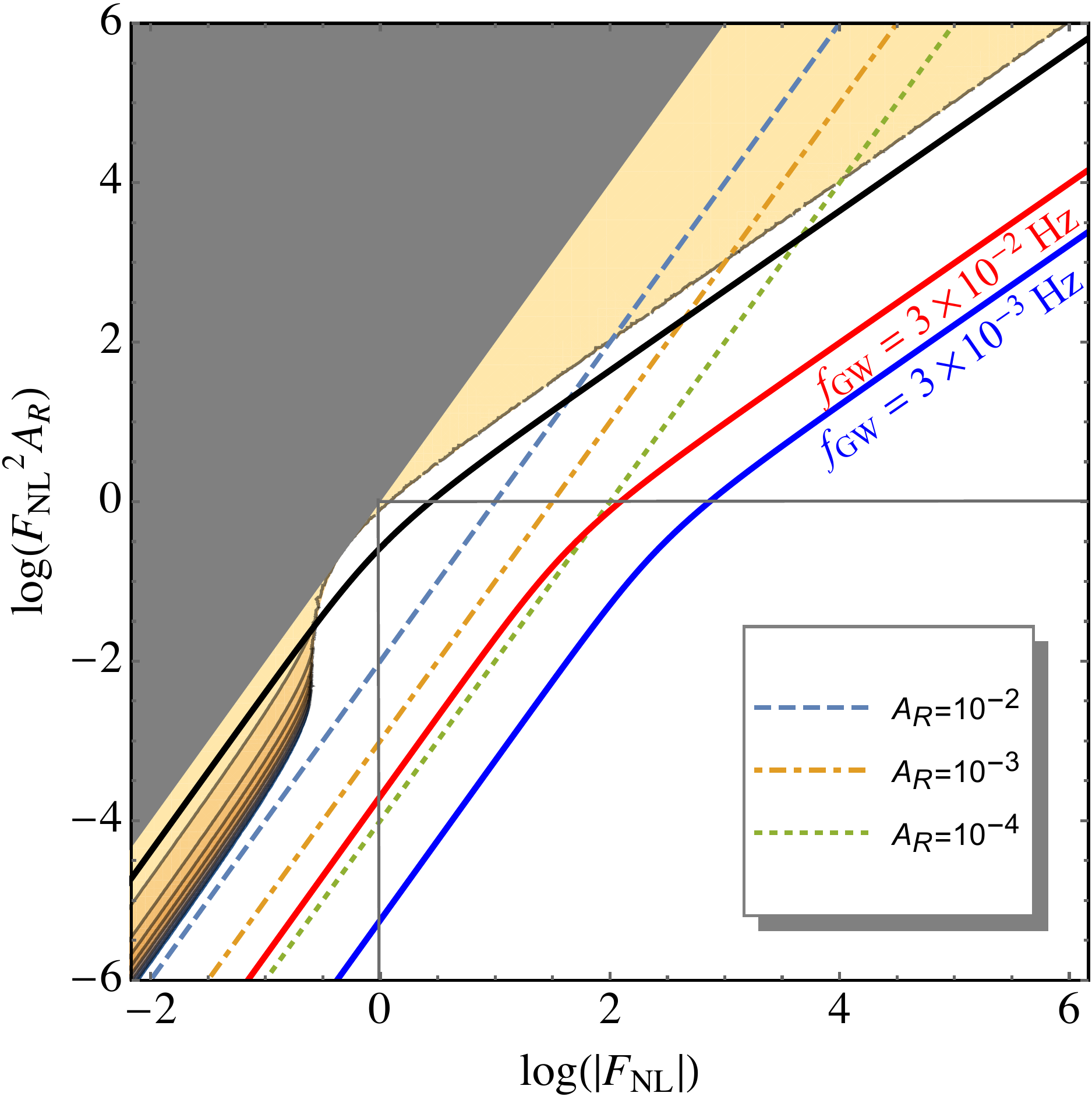}\includegraphics[width=0.05\textwidth]{legends.pdf}
	\caption{The primordial black hole mass fraction at formation $\beta$ depicted as a function of $\fnl$ and $\fnl^2\mathcal{A}_\mathcal{R}$, for the positive $\fnl$ (up) and the negative $\fnl$ (down), respectively. The constant $\beta$ contours are drawn, where the upper bound given by $\beta<\{8.6\times10^{-16},8.6\times10^{-15}\}$ for the PBHs corresponding to PBH masses at $M_\text{PBH}=\{10^{20}~\text{g},10^{22}~\text g\}$ can be seen as the border of the white and colored areas. The dashed lines are for $\mathcal{A}_\mathcal{R}=10^{-2}$, $10^{-3}$, and $10^{-4}$ from left to right, while the shaded area is unphysical since $\mathcal{A}_\mathcal{R}>1$. The thick black curve is the absolute constraint that the GW energy density be smaller than the current density of radiation, while the red and blue curves are the sensitivity bound of LISA at $f_\text{GW}=3\times10^{-2}~\text{Hz}$ and $3\times10^{-3}~\text{Hz}$, respectively; they correspond to PBH masses $M_\text{PBH}=10^{20}~\text g$ and $10^{22}~\text g$.}
\label{fig:beta}
\end{figure}

We can also constrain $\fnl$ on small scales by the abundance of PBHs. For the non-Gaussian curvature perturbation, 
Eq. \eqref{def:NGR}, the tadpole term  in our case is given by 
$\left\langle\mathcal{R}_g^2(\mathbf{x})\right\rangle=\int d^3kP_\mathcal{R}=\mathcal{A}_\mathcal{R}$.
Then we can express the Gaussian perturbation $\mathcal{R}_g$ in terms of $\mathcal{R}$ 
as in Ref.~\cite{Young:2013oia}, 
\be\nn
\mathcal{R}_{g\pm}(\mathcal{R})
=\frac12\fnl^{-1}\left(-1\pm\sqrt{1+4\fnl\left(\fnl\mathcal{A}_\mathcal{R}+\mathcal{R}\right)}\right).
\ee
PBHs will form if the curvature perturbation exceeds some threshold value $\mathcal{R}_\text{th}\sim1$~\cite{Musco:2004ak,Musco:2008hv,Musco:2012au,Harada:2013epa}.
The PBH mass fraction at the formation is
\be
\beta\simeq\left\{
\begin{matrix}
\frac12\text{erfc}\left(\frac{\mathcal{R}_{g+}(\mathcal{R}_\text{th})}{\sqrt{2\mathcal{A}_\mathcal{R}}}\right)
-\frac12\text{erfc}\left(-\frac{\mathcal{R}_{g-}(\mathcal{R}_\text{th})}{\sqrt{2\mathcal{A}_\mathcal{R}}}\right);&\fnl>0,\\\\
\frac12\text{erf}\left(\frac{\mathcal{R}_{g+}(\mathcal{R}_\text{th})}{\sqrt{2\mathcal{A}_\mathcal{R}}}\right)
-\frac12\text{erf}\left(\frac{\mathcal{R}_{g-}(\mathcal{R}_\text{th})}{\sqrt{2\mathcal{A}_\mathcal{R}}}\right);&\fnl<0.
\end{matrix}
\right.
\ee
For definiteness, we assume that the curvature perturbation peaks at $3\times10^{-3}~\text{Hz}$, which generates PBHs
 with a single mass of $10^{22}~\text{g}$. 
 There are basically no observational constraints on the PBH abundance for the mass range $10^{20}\sim10^{22}~\text{g}$~\cite{Niikura:2017zjd} and $10^{17}\sim10^{19}~\text{g}$~\cite{Barnacka:2012bm,Katz:2018zrn}, 
 except for the constraint that the PBH density cannot exceed that of dark matter, i.e. 
 $1\geq \Omega_\text{PBH}/\Omega_\text{DM}\approx1.16\times10^{17}\beta(M_\text{PBH}/10^{16}~\text{g})^{-1/2}$. 
  This relation together with \eqref{frequency-mass} gives
\be\label{betaconstraint}
\beta\lesssim8.6\times10^{-15}\left(\frac{3\times10^{-3}~\text{Hz}}{f_\text{GW}}\right).
\ee
Constraint \eqref{betaconstraint} is drawn in Fig.\ref{fig:beta} for both positive and negative $\fnl$, together with the sensitivity bound of LISA from $f_\text{GW}=3\times10^{-2}$ to $3\times10^{-3}~\text{Hz}$.
The white area in both figures is the parameter space allowed. 
When $\fnl\lesssim -0.3$, it is impossible to generate enough PBHs to account for dark matter 
since there would be too much GWs, which means there is no constraint from PBHs when $\fnl$ is negative.

For $\fnl>0$, the parameter space is narrower. From the small $\fnl$ limit, we see that to avoid PBH overproduction 
we need $\mathcal{A}_\mathcal{R}\lesssim1.5\times10^{-2}$. Besides, for a given $\mathcal{A}_\mathcal{R}$,
 there is an upper bound for $\fnl$ from the PBH abundance constraint \eqref{betaconstraint}:
  $\fnl<0.017/\mathcal{A}_\mathcal{R}$. This can be found from the intersections of the PBH constraint and the
   equal-$\mathcal{A}_\mathcal{R}$ lines in Fig. \ref{fig:beta}. Interestingly, all of the possible PBH abundances 
   are above the LISA sensitivity curve, which means that if PBHs with masses from $10^{20}$ to $10^{22}~\text{g}$ are the dominant dark matter,
  we must observe the corresponding GW signals by LISA, no matter how small $\mathcal{A}_\mathcal{R}$ is.

\textit{Conclusion}~
We studied the effect of a local-type non-Gaussianity 
in the curvature perturbation on the induced
tensor perturbation at second order as well as on the PBH formation.
The scalar perturbation was assumed to have a narrow peak on a small scale $1/k_*$, with a local-type non-Gaussianity.
Our result shows that if $\mathcal{A}_\mathcal{R}\fnl^2\gtrsim1$,  the non-Gaussian contribution 
becomes prominent,
and the main features of the GW density parameter $\Omega_\text{GW}$ will be a series of peaks with 
the highest at $(4/\sqrt3)k_*$ just before the cutoff at $4k_*$, and the $k^3$ slope on the smaller $k$ side of the peaks. The detection of these features will be clear evidence for the primordial non-Gaussianity of the curvature perturbation at around $k_*$.

In this Letter we only considered a narrow peak in the scalar perturbation spectrum, although broad plateaus may be generated in some other models of inflation~\cite{GarciaBellido:1996qt,Kawasaki:1997ju,Yokoyama:1998pt,Kohri:2012yw,Clesse:2015wea,Inomata:2017okj,Garcia-Bellido:2017mdw,Kannike:2017bxn,Inomata:2017vxo,Ando:2017veq,Ando:2018nge,Espinosa:2017sgp,Espinosa:2018eve,Cheng:2016qzb,Cheng:2018yyr,Garcia-Bellido:2016dkw}. Nevertheless, our criterion for the existence of non-Gaussianity remains universal. We can see from the integral \eqref{OmegaFull} that the power $\beta$ of $\Omega_\text{GW}\sim k^\beta$ induced by the Gaussian scalar perturbations will be around 3 when $k\ll\sigma\ll k_*$, but decreases as the width $\sigma$ increases, while $\sigma\rightarrow\infty$ will induce a scale-invariant GW spectrum as expected, which is also shown numerically in Ref. \cite{Bugaev:2009zh}. So we can conclude that $\beta\lesssim3$ is characteristic for GWs induced by scalar perturbations.  The first order electroweak phase transition may also give rise to stochastic GWs with $\beta\sim3$ on the low frequency side~\cite{Caprini:2009fx,Jinno:2016vai}. However, almost all of the previous results indicate that the peak frequency is below the LISA band, thus we can probably only detect the high frequency tail where $\beta<0$ by LISA~\cite{Kosowsky:1992vn,Kamionkowski:1993fg,Apreda:2001us,Grojean:2006bp,Caprini:2007xq,Caprini:2009yp,Hindmarsh:2013xza,Cai:2017tmh,Chao:2017vrq,Huang:2016odd}. Another possible source 
is the incoherent superpositions of GWs from compact binaries, 
which has 
$\beta\sim2/3$~\cite{Zhu:2011bd}. This means that the detection of GWs with $\beta\sim3$ can be recognized as of induced origin, where multiple peaks will be a smoking gun of primordial non-Gaussianity. Further detailed studies are left for future work. 

We also derived constraints on the PBH abundances. 
Currently it is possible for PBHs to serve as all the dark matter if $M_\text{PBH}$ locates 
in the range $10^{17}$ to $10^{19}~\text{g}$ or $10^{20}$ to $10^{22}~\text{g}$. The former case corresponds to GWs with peak frequency from 0.1 to 1~Hz, which can be fully explored by DECIGO, while the low frequency tail can be seen by LISA. 
In this Letter we focus on the latter case which corresponds to the GW frequencies $3\times10^{-3}$ to $3\times10^{-2}~\text{Hz}$, right 
in the sensitivity frequency band of LISA. 
We found that if these PBHs consist a substantial portion of
 the dark matter, the corresponding GW signal must be detectable by LISA. Conversely, if we are unable to detect any induced GW signal by LISA, it will be impossible for PBHs to serve as all dark matter in the mass range $10^{20}$g to $10^{22}$g. Depending on the integration time, the abundance of PBHs can be further constrained. Therefore the induced GWs can be used as a powerful tool of probing the abundances of small PBHs. This will also be left for our future work.

\textit{Acknowledgements}~
We thank Bin Hu, Ryo Saito, Lijing Shao, Masahiro Takada, and Yi-Peng Wu for useful discussions. S.P. thanks the Institute for Theoretical Physics, CAS for the hospitality during his visit. R.G.C. was supported by the National Natural Science Foundation of China Grants No.~11435006, No. 11647601, No. 11690022, and No. 11851302, and by the Strategic Priority Research Program of CAS Grant No.~XDB23030100, and by the Key Research Program of Frontier Sciences of CAS. S.P. and M.S. were supported by the MEXT/JSPS KAKENHI No.~15H05888 and No. 15K21733, and by the World Premier International Research Center Initiative (WPI Initiative), MEXT, Japan.

\end{document}